\documentclass[prl,twocolumn,usletter,showpacs,superscriptaddress]{revtex4}

\usepackage{amsmath}
\usepackage{amsfonts}
\usepackage{amssymb}

\usepackage [ansinew] {inputenc}
\usepackage{graphicx}
\usepackage{natbib}

\begin{document}
\title{Cellular automaton for chimera states}

\author{Vladimir Garc\'{\i}a-Morales}
\email{garmovla@uv.es}
\affiliation{Departament de Termodin\`amica, Universitat de Val\`encia, E-46100 Burjassot, Spain}

\begin{abstract}
\noindent A minimalistic model for chimera states is presented. The model is a cellular automaton (CA) which depends on only one adjustable parameter, the range of the nonlocal coupling, and is built from elementary cellular automata and the majority (voting) rule. This suggests the universality of chimera-like behavior from a new point of view: Already simple CA rules based on the majority rule exhibit this behavior. After a short transient, we find chimera states for arbitrary initial conditions, the system spontaneously splitting into stable domains separated by static boundaries, ones synchronously oscillating and the others incoherent. When the coupling range is local, nontrivial coherent structures with different periodicities are formed.
\end{abstract}

 \maketitle \pagebreak
\maketitle

%\section{Section title }

Chimera states arise in sets of identical oscillators as a result of their stable grouping into two separated subsets, one of them synchronously oscillating, the other incoherent. This phenomenon was first pointed out in a network of oscillators under a symmetric nonlocal coupling \cite{ChimeraKUR02a,ChimeraABR04}. Chimera states were then experimentally discovered in populations of coupled chemical oscillators \cite{ChimeraTIN12} and in optical coupled-map lattices realized by liquid-crystal light modulators \cite{ChimeraHAG12}. Great theoretical \cite{ChimeraPAN15,ChimeraABR08,ChimeraSET08,ChimeraLAI09,ChimeraMOT10,ChimeraMAR10,ChimeraOLM10,ChimeraBOR10,ChimeraSHE10,ChimeraWOL11,ChimeraLAI11,ChimeraOME11,ChimeraOME12,ChimeraOME13,ChimeraNKO13,ChimeraHIZ13,ChimeraSET13,VGM3,ChimeraSET14,ChimeraYEL14,ChimeraBOE15,ChimeraBUS15,ChimeraOME15,ChimeraOME15A, ChimeraBAN15, ChimeraASH14, ChimeraASHW15b, ChimeraHIZ15, ChimeraSCHM15a, ChimeraSCHM15, ChimeraHAU15, ChimeraKEM16} and experimental \cite{ChimeraMAR13,ChimeraLAR13,ChimeraKAP14,ChimeraWIC13,ChimeraWIC14,ChimeraSCH14a,ChimeraGAM14,ChimeraROS14a,ChimeraLAR15} interest followed. Chimera states may also describe some aspects of the dynamical behavior of social systems \cite{ChimeraGON14}, power grids \cite{ChimeraMOT13a}, epileptic seizures \cite{ChimeraROT14} and the unihemispheric sleep of birds and dolphins \cite{ChimeraRAT00}.
This motivates the need of simple models, reduced to the barest essentials, to describe the underlying mechanisms behind their formation. Cellular automata (CAs) \cite{Wolfram, Ilachinski, Adamatzky, McIntosh, Wuensche, Ceccherini, VGM1, VGM2, VGM3} hold promise for that goal. For example, chimera states were found in a three-level CA of Zykov-Mikhailov type \cite{ChimeraMAK}, and Boolean phase oscillators, realized with electronic logic circuits, were also shown to support transient chimeras \cite{ChimeraROS14a}.

In this letter we regard chimera states as an experimental fact of nature rather than a feature of certain systems of differential equations or maps. We then formulate a simple CA model for chimera states, describing a possible universal mechanism behind their spontaneous emergence out of any initial condition. We sketch a general mathematical approach to show how CAs can be regarded as approximations (shadowings) of maps of coupled oscillators. Although we do not attempt here to connect our specific CA model to any such  map, we hypothesize that the latter should exist \cite{JPHYSA, Omohundro}. Chimeras are here modelled as specific instances of domain formation in spatially extended systems, a behavior that is statistically robust to small perturbations and which is ubiquitously found in nature. The chimera states encountered here are of the weak type \cite{ChimeraASH14, ChimeraASHW15b} and are \emph{stable} and coexist with synchronously oscillating domains separated by static walls. The model depends on one free parameter only, $\xi \in \mathbb{N}$ $(\xi \ge 1)$, whose physical meaning is the neighborhood radius (nonlocal coupling range). When $\xi$ is small, nontrivial coherent structures are formed. However, when $\xi$ is sufficiently large, incoherent domains of thickness $d \ge \xi+1$ arise.  

We first show how any map on a ring of $N$ spatially coupled oscillators can be approximated by a CA.  Let $\varphi_{t}^{j} \in [0, 2\pi)$ denote the phase of the oscillator at location $j$, $j \in [0,N-1]$ and discrete time $t$. We assume that the evolution of the phases in the torus $T^{N}$ is governed by   
\begin{equation}
\varphi_{t+1}^{j}=F\left(\varphi_{t}^{j+\xi}, \varphi_{t}^{j+\xi-1}, \ldots, \varphi_{t}^{j}, \ldots, \varphi_{t}^{j+1-\xi}, \varphi_{t}^{j-\xi}\right) \label{phases}
\end{equation}
where $F: T^{N}\to [0,2\pi)$ is a continuous nonlinear function that couples the oscillators within a range $\xi$. We assume that the oscillators are identical (same natural frequency $\omega$) and indistinguishable, i.e. that Eq. (\ref{phases}) is invariant, modulo $N$, to an arbitrary permutation of the labels \cite{ChimeraASH14}. A specific instance is considered in, e.g., \cite{ChimeraOME11}.  

A rigorous definition of a chimera state has been recently given in \cite{ChimeraASH14, ChimeraASHW15b}. Oscillators $i$ and $j$ are frequency synchronized if $\Omega_{ij}\equiv \lim_{t\to \infty} \left(\varphi_{t}^{j}-\varphi_{t}^{i}\right)/t=0$ \cite{ChimeraASH14}.  A flow-invariant $\omega$-limit set on the torus $T^{N}$, $[0,2\pi)^{N}$  is a weak chimera state if there exist three oscillators $i$, $j$ and $k$ such that $\Omega_{ij}\ne 0$ and $\Omega_{ik}= 0$ \cite{ChimeraASH14}. In this work we provide a construction that is shown to support weak chimera states. We first explain how to (approximately) map the dynamics on the torus $T^{N}$ to the shift space $\mathcal{A}^{N}$ \cite{Lind} of a CA. Here $\mathcal{A}$ denotes the set of integers in $[0,p-1]$ with $p\in \mathbb{N}$ $(p\ge 2)$ being the alphabet size. 

Since $\frac{\varphi}{2\pi} \in [0,1)$ is a real number, we can expand it in a base (radix) $p\ge 2$, $p \in \mathbb{N}$ as
\begin{equation}
\frac{\varphi_{t}^{j}}{2\pi}=\lim_{D\to \infty} \sum_{m=1}^{D}p^{-m}\mathbf{d}_{p}\left(-m,\frac{\varphi_{t}^{j}}{2\pi}\right) \label{phasexpan}
\end{equation}
where we have introduced the digit function \cite{CHAOSOLFRAC, PHYSAFRAC, semipredo}
\begin{equation}
\mathbf{d}_{p}(k,x)=\left \lfloor x/p^{k} \right \rfloor-p\left \lfloor x/p^{k+1} \right \rfloor  \label{cucuA}
\end{equation}
for any $x \in \mathbb{R}$, $k \in \mathbb{Z}$. Here $\left \lfloor \ldots \right \rfloor$ denotes the floor (lower closest integer) function and we have $0 \le \mathbf{d}_{p}(k,x) \le p-1$. 
If we now expand in radix $p$ both sides of Eq. (\ref{phases}), terms with same powers of $p$ are equal because the radix-$p$ representation is \emph{unique} for any rational number arising from truncating a real number to finite $D$ \cite{Andrews}. From Eq. (\ref{phases}) we thus have a set of equations
\begin{equation}
\mathbf{d}_{p}\left(-m,\frac{1}{2\pi}\varphi_{t+1}^{j}\right)=\mathbf{d}_{p}\left(-m,\frac{1}{2\pi}F\left(\varphi_{t}^{j+\xi}, \ldots, \varphi_{t}^{j-\xi}\right)\right) \nonumber
\end{equation}
where $m \in [1, D]$, $m\in \mathbb{Z}$. A CA approximation of Eq. (\ref{phases}) is obtained by considering only the dynamical behavior of the first digit after the radix point $(m=1)$ of the phases. If we then take $\varphi_{t}^{j} \approx 2\pi \mathbf{d}_{p}\left(-1,\frac{\varphi_{t}^{j}}{2\pi}\right)/p$ and define 
\begin{eqnarray}
x_{t}^{j}&\equiv& \mathbf{d}_{p}\left(-1,\frac{1}{2\pi}\varphi_{t}^{j}\right) \label{xtj} \\
f\left(x_{t}^{j+\xi}, \ldots, x_{t}^{j-\xi}\right) &\equiv& \mathbf{d}_{p}\left(-1,\frac{1}{2\pi}F\left(\varphi_{t}^{j+\xi}, \ldots, \varphi_{t}^{j-\xi}\right)\right) \nonumber
\end{eqnarray}
we obtain a CA dynamics
\begin{equation}
x_{t+1}^{j}=f\left(x_{t}^{j+\xi}, x_{t}^{j+\xi-1}, \ldots, x_{t}^{j}, \ldots, x_{t}^{j+1-\xi}, x_{t}^{j-\xi}\right) \label{CAdyna}
\end{equation}
with $x_{t}^{j}\in [0,p-1]$. The above approximation becomes more accurate as $p$ is increased. Since $x_{t}^{j}=\mathbf{d}_{p}(0,x_{t}^{j})$, if we take, e.g., $p=8$ we have 
\begin{eqnarray}
\mathbf{d}_{8}(0,x_{t}^{j})&=&x_{t}^{j}-8\left \lfloor x_{t}^{j}/8 \right \rfloor = x_{t}^{j}-2\left \lfloor x_{t}^{j}/2 \right \rfloor+2\left \lfloor x_{t}^{j}/2 \right \rfloor- \nonumber \\ && -4\left \lfloor x_{t}^{j}/4 \right \rfloor +4\left \lfloor x_{t}^{j}/4 \right \rfloor-8\left \lfloor x_{t}^{j}/8 \right \rfloor \nonumber \\
&=&\mathbf{d}_{2}(0,x_{t}^{j})+2\mathbf{d}_{2}\left(0,x_{t}^{j}/2\right)+4\mathbf{d}_{2}\left(0,x_{t}^{j}/4\right) \nonumber 
\end{eqnarray}
Hence, if we define
\begin{equation}
y^{(h),j}_{t} \equiv \mathbf{d}_{2}\left(0, x_{t}^{j}/2^{h} \right) \qquad h=0,1,2 \label{ini0}
\end{equation}
we observe that at $t$ and $j$ we can write $x_{t}^{j}=y_{t}^{(0),j}+2y_{t}^{(1),j}+4y_{t}^{(2),j}$
where each $y_{t}^{(h),j}$ $(h=0,1,2)$ is either zero or one. We shall call the specific value of  $h$ the \emph{layer} of the CA. At $t+1$ we have, similarly 
\begin{equation}
x_{t+1}^{j}=y_{t+1}^{(0),j}+2y_{t+1}^{(1),j}+4y_{t+1}^{(2),j} \label{modelatp1}
\end{equation}
Thus, the local transformation $x_{t}^{j}\to x_{t+1}^{j}$ on 8 symbols is equivalent to a local transformation $(y^{(0),j}_{t},y^{(1),j}_{t},y^{(2),j}_{t}) \to (y^{(0),j}_{t+1},y^{(1),j}_{t+1},y^{(2),j}_{t+1})$ on triples of Boolean variables. From Eqs. (\ref{CAdyna}) and (\ref{ini0}), we have 
\begin{equation}
y^{(h),j}_{t+1} \equiv \mathbf{d}_{2}\left(0, \frac{1}{2^{h}}f\left(x_{t}^{j+\xi}, \ldots, x_{t}^{j-\xi}\right) \right) \label{coupling}
\end{equation}
so that, in general, all layers $h$ are nonlinearly coupled within the neighborhood of radius $\xi$. 

Our guiding principle now is \emph{to identify at the CA level a nonlocal coupling among the $x_{t}^{j}$'s that leads the oscillators to split into two groups (clustering) \emph{and} that is also able to adopt a different form on each group.} A most simple way of achieving this is, e.g., to make the coupling of the layers $h$ entirely dependent on the value $y_{t}^{(0),j}$ of layer $h=0$ only. Thus, when $y_{t}^{(0),j}=0$ ($x_{t}^{j}$ even) let the coupling be \emph{synchronizing} and when $y_{t}^{(0),j}=1$ ($x_{t}^{j}$ odd), let it be \emph{desynchronizing}.  We now formulate our CA model for chimera-like behavior. Although we discuss the model at the CA level only, we hypothesize that there should exist a coupled map lattice from which the model is an approximation \cite{JPHYSA}. Let $x_{t}^{j} \in [0,7]$. Then at time $t+1$ 
\begin{equation}
x_{t+1}^{j}=f_{t}^{(0),j}+2\left(1-f_{t}^{(0),j}\right)f_{t}^{(1),j}+4f_{t}^{(0),j}f_{t}^{(2),j} \label{themodel}
\end{equation}
where the $f_{t}^{(h),j}\in \{0,1\}$ $(h=0,1,2)$ are given by
\begin{eqnarray}
\label{layers}
f_{t}^{(0),j}&\equiv&H\left(-\frac{1}{2}+\frac{1}{2\xi+1}\sum_{k=-\xi}^{\xi}y_{t}^{(0),j+k} \right) \label{majomo}\\
f_{t}^{(1),j}&\equiv&1-H\left(-\frac{1}{2}+\frac{1}{2\xi+1}\sum_{k=-\xi}^{\xi}y_{t}^{(1),j+k} \right) \label{oscimo}\\
f_{t}^{(2),j}&\equiv&\mathbf{d}_{2}\left(0,1+y_{t}^{(2),j+1}+y_{t}^{(2),j}+y_{t}^{(2),j-1} \right) 
\label{ado}
\end{eqnarray}
Here the $y_{t}^{(h),j}$'s are obtained from Eqs. (\ref{ini0}) and $H(x)$ is the Heaviside function ($H(x)=0$ for $x <0$, $H(0)=\frac{1}{2}$ and $H(x)=1$ for $x >0$). The model evolves as follows. From an initial condition $x_{0}^{j}$, specified at every $j \in [0,N-1]$, the $y_{0}^{(h),j}$'s are calculated from Eq. (\ref{ini0}). Then, they are inserted in Eqs. (\ref{majomo}) to (\ref{ado}) so that the $f_{t}^{(h),j}$'s are obtained. By replacing them in Eq. (\ref{themodel}), $x_{1}^{j}$ is calculated. This process is iterated $t$ times to yield $x_{t}^{j}$, $\forall j$.

\begin{figure} %Shown is a window where $t \in [0,150]$ and $i\in [1, 150]$
\includegraphics[width=0.49 \textwidth]{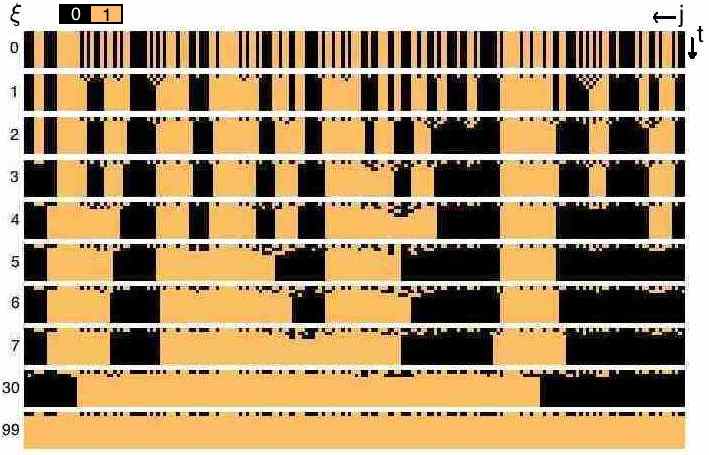}
\caption{\scriptsize{Ten iterates of the majority rule, Eq. (\ref{majo}), for a ring of $N=200$ sites starting from a random initial condition (which is the same for all panels) for the values of $\xi$ indicated in the figure.}} \label{majori}
\end{figure}

By comparing Eqs. (\ref{modelatp1}) and (\ref{themodel}) we observe that
\begin{eqnarray}
y_{t+1}^{(0),j}&=&\mathbf{d}_{2}(0,x_{t+1}^{j})=f_{t}^{(0),j} \label{layer0} \\
y_{t+1}^{(1),j}&=&\mathbf{d}_{2}\left(0,x_{t+1}^{j}/2\right)=\left(1-f_{t}^{(0),j}\right)f_{t}^{(1),j} \label{layer1} \\
y_{t+1}^{(2),j}&=&\mathbf{d}_{2}\left(0,x_{t+1}^{j}/4\right)=f_{t}^{(0),j}f_{t}^{(2),j} \label{layer2}
\end{eqnarray}
These equations specify the couplings within layers of the CA in Eq. (\ref{themodel}). Layer $h=0$ is decoupled from layers $h=1$ and $h=2$ at every $t$ and $j$ but it influences those layers. The spatiotemporal behavior of the layer $h=0$ is thus, independent of the other layers, and is dictated by Eqs. (\ref{majomo}) and (\ref{layer0}) as
\begin{equation}
\label{majo}
y_{t+1}^{(0),j}=H\left(-\frac{1}{2}+\frac{1}{2\xi+1}\sum_{k=-\xi}^{\xi}y_{t}^{(0),j+k} \right) 
\end{equation}
This is the majority (voting) rule \cite{Vichniac, Tchuente, Goles, GolesBOOK, Ilachinski}. The  evolution of Eq. (\ref{majo}) for a random initial condition of zeros and ones is shown in Fig. \ref{majori} for the values of $\xi$ indicated. It is well-known that the majority rule has stable spatial fixed points \cite{Tchuente, Goles, GolesBOOK, Ilachinski}. Indeed, Agur \cite{Agur} found that the number of such stable fixed points $\mathcal{F}(N, \xi)$ is given by
\begin{equation}
\mathcal{F}(N, \xi)=2+\sum_{\ell=1}^{\left \lfloor \frac{N}{2(\xi+1)} \right \rfloor}\frac{2N}{N-2\ell \xi}{N-2\ell \xi \choose 2\ell } \label{AgurAgur}
\end{equation}
We see that $\mathcal{F}(N, \xi)$ decreases by increasing $\xi$ for fixed $N$. She also showed  that the thickness $d$ of the spatial domains satisfy $d \ge \xi+1$ so that $\xi$ is, indeed, \emph{a rigorous lower bound} for $d$ \cite{Agur}. These facts are all observed in Fig. \ref{majori}: After a short transient the system converges to a spatial fixed point were the size of the domains is larger for $\xi$ large. If $\xi=\left \lfloor N/2 \right \rfloor$ with $N$ odd (i.e. if $N=2\xi+1$), the neighborhood of site $j$ coincides with the whole ring in which case we have a \emph{global coupling}. There are only two fixed points  in this case (all sites '0' or all sites '1'), as it is simply obtained from Eq. (\ref{AgurAgur}), since $\mathcal{F}(N, \left \lfloor N/2 \right \rfloor)=2$. A useful measure of the robustness of the fixed points is the system's \emph{resilience} $\mathcal{R}(N,\xi)$ given by \cite{Agur,Agur2}
\begin{equation}
\mathcal{R}(N,\xi)=1-\frac{\mathcal{F}(N, \xi)}{2^{N}} \label{resina}
\end{equation}
This quantity measures the probability that a fixed point remains unaltered if a single bit is changed \cite{Agur}. We see that $\mathcal{R}(N,\xi)$ is larger for $\xi$ large, which means that the domains are more robust to perturbations as $\xi$ is increased. A bound for the duration of the transient $\tau$ in the majority rule is also known \cite{GolesBOOK, Ilachinski}. Let $d_{0}$ be the maximum thickness of any finite block of nonzero sites within the initial condition. Then \cite{GolesBOOK, Tchuente} 
\begin{equation}
\tau \le (\xi+d_{0}+2)\xi \label{transient}
\end{equation}
%Thus, any fixed point is reached for $t > \tau$.

%Let $N_{1}$ be the total number of oscillators with initial value $y_{0}^{(0),j}=1$ and $N_{0}$ the ones with initial value $y_{0}^{(0),j}=0$  so that $N=N_{0}+N_{1}$. Under global coupling, trivially, there is a single domain, with all sites equal to one if the initial density $\rho=N_{1}/N>0.5$ or all sites equal to zero if $\rho < 0.5$ (note that $\rho=0.5$ is not possible for $N$ an odd number). 

\begin{figure} %Shown is a window where $t \in [0,150]$ and $i\in [1, 150]$
\includegraphics[width=0.49 \textwidth]{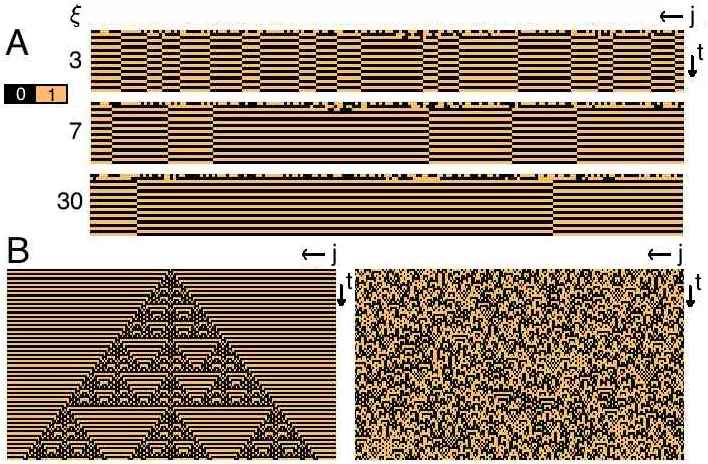}
\caption{\scriptsize{A. Spatiotemporal evolution of Eq. (\ref{oscirule}), for  values of $\xi$ indicated in the figure $N=200$ and 20 time steps. The initial condition is the same as in Fig. (\ref{majori}). B. Spatiotemporal evolution of Wolfram's rule 105, Eq. (\ref{W105}) for $N=200$, $95$ time steps and for a simple initial condition of a single site with value '1' surrounded by zeroes (left panel) and a generic initial condition (right panel) that is the same as in Fig. \ref{majori}.}} \label{oscianadi}
\end{figure} 

\begin{figure*} %Shown is a window where $t \in [0,150]$ and $i\in [1, 150]$
\includegraphics[width=1.0 \textwidth]{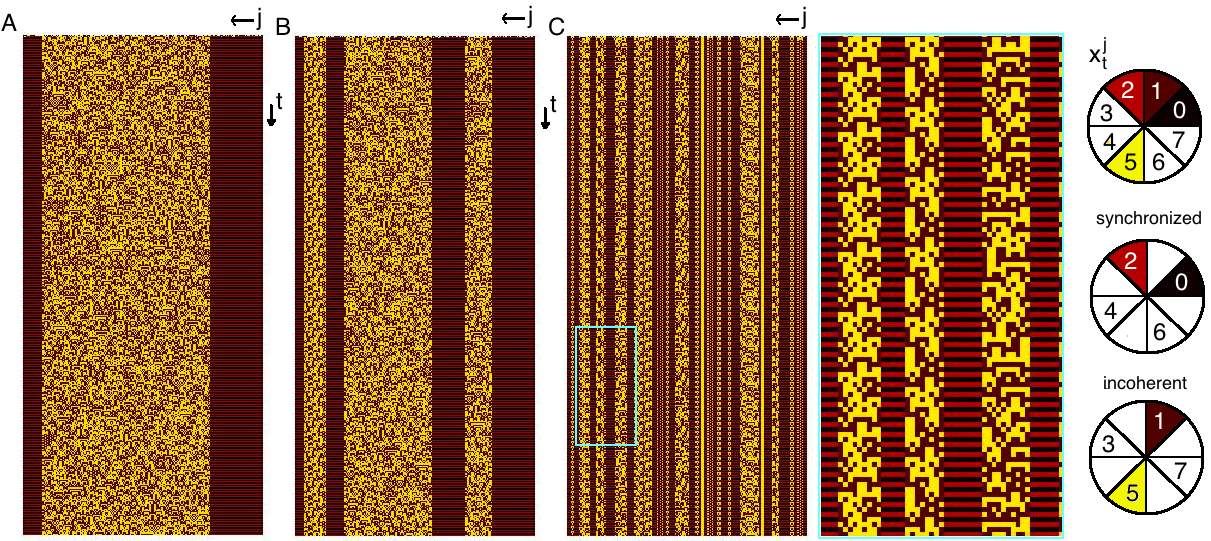}
\caption{\scriptsize{Spatiotemporal evolution of $x_{t}^{j}$ provided by Eq. (\ref{themodel}) for $\xi=30$ (A), $7$ (B) and $1$ (C), for $N=200$ and $400$ time steps and for an initial condition $x_{0}^{j}=a^{j}+2a^{j}+4a^{j}$, where $a^{j} \in \{0,1\}$ is as in Fig. (\ref{majori}). The rightmost panel is a detail of panel C. The color code for the site values corresponding to synchronized and incoherent domains is also shown. The values $x_{t}^{j}=3,4,6,7$ do not occur in the trajectory.}} \label{chimepl}
\end{figure*}

In domains where $y_{t}^{(0),j}=0$, layer $h=2$ is in the quiescent state ($y_{t}^{(2),j}=0$, from Eq. (\ref{layer2})). In layer $h=1$, $y_{t+1}^{(1),j}=f_{t}^{(1),j}$ from Eq. (\ref{layer1}), and Eq. (\ref{oscimo}) reduces to 
\begin{equation}
\label{oscirule}
y_{t+1}^{(1),j}=1-H\left(-\frac{1}{2}+\frac{1}{2\xi+1}\sum_{k=-\xi}^{\xi}y_{t}^{(1),j+k} \right)
\end{equation}
This CA is similar to the majority rule above, but generates oscillations between values '0' and '1', having no fixed-points. It is trivial to show that Eq. (\ref{oscirule}) has a 2-cycle once one has shown that the majority rule Eq. (\ref{majo}) has a fixed point. For, by noting that $H(-x)=1-H(x)$, and iterating Eq. (\ref{oscirule}) twice, we find $x_{t+2}=1-H\left(\frac{1}{2}-\frac{1}{2\xi+1}\sum_{k=-\xi}^{\xi}H\left(-\frac{1}{2}+\frac{1}{2\xi+1}\sum_{k'=-\xi}^{\xi}x_{t}^{j+k+k'} \right) \right)=H\left(-\frac{1}{2}+\frac{1}{2\xi+1}\sum_{k=-\xi}^{\xi}H\left(-\frac{1}{2}+\frac{1}{2\xi+1}\sum_{k'=-\xi}^{\xi}x_{t}^{j+k+k'} \right) \right)$ which is equal to two iterates of Eq. (\ref{majo}). Thus, at the fixed point of Eq. (\ref{majo}), we have a 2-cycle of Eq. (\ref{oscirule}).
In Fig. \ref{oscianadi}A the spatiotemporal evolution of this rule is shown for several different values of $\xi$. Domains are formed as in the majority rule case, (compare with Fig. \ref{majori}) but each individual site instead of being at a fixed point, synchronously oscillates in phase with all sites within its same domain. Eq. (\ref{W105}) can be considered as a toy model for phase clusters in absence of phase balance, as it was described for the Belousov-Zhabotinsky reaction under global feedback \cite{Dolnik} and for electrochemical systems under galvanostatic constraint \cite{KrischerBaba}.

In domains where $y_{t}^{(0),j}=1$ we have, from Eq. (\ref{layer1}), that layer $h=1$ is quiescent ($y_{t}^{(1),j}=0$) and layer $h=2$ is active. We have $y_{t+1}^{(2),j}=f_{t}^{(2),j}$ from Eq. (\ref{layer2}) and, hence, Eq. (\ref{ado}) becomes 
\begin{equation}
\label{W105}
y_{t+1}^{(2),j}=\mathbf{d}_{2}\left(0,1+y_{t}^{(2),j+1}+y_{t}^{(2),j}+y_{t}^{(2),j-1} \right)
\end{equation}
This is Wolfram CA rule 105. It has positive left $\lambda_{L}=1$ and right $\lambda_{R}=1$ Lyapunov exponents (see Table 6, p. 541 in \cite{Wolfram2}) and, thus, it rigorously qualifies as a \emph{chaotic} CA. Rules of this kind were considered in pioneering works on spatiotemporal intermittency \cite{Chate2, Chate3} and their triangular structures strikingly resemble those encountered in the complex Ginzburg-Landau equation \cite{KuramotoBOOK, Aranson, contemphys} in the regime of spatiotemporal chaos. Furthermore, rule 105 is a totalistic additive CA rule that depends only on the sum over neighborhood values and, hence, it is directly related to a discretized version of the Laplacian (diffusion) operator. The rule has also a homogeneous 2-cycle as possible solution. However, for a generic initial condition, the spatiotemporal evolution of the rule is incoherent. In Fig. \ref{oscianadi} B we show $y_{t}^{(2),j}$ obtained from Eq. (\ref{W105}) for a simple initial condition consisting of a single site with value '1' surrounded by zeroes (left panel) and an arbitrary initial condition (right panel) that is the same as in Fig. \ref{majori}. Although a nested regular pattern is observed in the former case,  incoherent behavior is found in the latter one. 

We now study the spatiotemporal evolution of the CA, Eq. (\ref{themodel}).
In Fig. \ref{chimepl}, $x_{t}^{j}$ as obtained from Eq. (\ref{themodel}), is shown for $\xi=30$ (A), $7$ (B) and $1$ (C), for $N=200$ sites and $400$ time steps and for a generic initial condition (see figure caption). After a short transient the system spontaneously splits into two different domains, one in which the oscillators synchronously oscillate ($x_{t}^{j}=0$ or $2$), with vanishing Lyapunov exponents $\lambda_{L}=\lambda_{R}=0$, and the other chaotic ($x_{t}^{j}=1$ or $5$), with $\lambda_{L}=\lambda_{R}=1$. The values $x_{t}^{j}=3,4,6,7$ do not occur in the trajectory of the CA and can only be present as initial conditions. 
For each oscillator $j$ and $t > \tau$ we define
\begin{equation}
a_{j}(t, \Delta t)\equiv \frac{1}{\Delta t}\sum_{n=t}^{t+\Delta t-1}\delta(x_{n}^{j}-x_{t}^{j}) \label{counts}
\end{equation}
which obeys $0 < a_{j}(t, \Delta t) \le 1$. Here $\delta(x)$, $x \in \mathbb{Z}$, is the Kronecker delta: $\delta(x)=1$ if $x=0$ and $\delta(x)=0$ otherwise. Thus, $a_{j}(t, \Delta t)\Delta t$ counts the number of instances in which the phase $x_{n}^{j}$ is equal to $x_{t}^{j}$ within the time interval $\Delta t \in \mathbb{N}$. In our setting, we say that oscillators $i$ and $j$ are frequency synchronized if $\Omega_{ij}\equiv a_{i}(t, \Delta t)-a_{j}(t, \Delta t)=0$, $\forall t > \tau$ and $\forall \Delta t \ge 1$. Two oscillators $i$ and $k$ are desynchronized if $\Omega_{ik} \ne 0$. The three oscillators $i$, $j$ and $k$ can be easily found and this proves that we have weak chimera states \cite{ChimeraASH14, ChimeraASHW15b}. The walls separating the domains are \emph{stable}, as are the patterns thus formed. The thickness $d$ of the domains is dictated by the majority rule on layer $h=0$, to which the whole dynamics is slaved. Hence, as explained above, $d \ge \xi+1$. From Eq. (\ref{resina}) we have that, for larger $d$, the incoherent domains are more robust to small perturbations. Because $d$ is bounded from below, the multiplicity $\mathcal{M}$ of incoherent domains is bounded from above by $\mathcal{M}(N, \xi) \le \left \lfloor \frac{N}{2(\xi+1)} \right \rfloor$. Therefore, \emph{for larger coupling range the number of incoherent domains is lower on the average}. Quite strikingly, when $\xi$ is small, e.g. $\xi=1$ as in panel C, complex coherent structures with well defined periodicity are observed. In the rightmost panel of Fig. \ref{chimepl} a detail of panel C is shown, where bands with thickness of $d=9, 7$ and $10$ sites contain structures with periods $T=12$, $14$ and $62$, respectively, all these coherent structures coexisting with the uniformly oscillating background of period $2$. In all cases, $d > \xi =1$. We note that, because of the finiteness of the dynamics, the period of any structure is bounded above by $T \le 2^{d}$, where the equality would only hold if the dynamics were ergodic (which is not). For the chimera state in panel A of Fig. \ref{chimepl} we would expect the pattern to be repeated before $\sim 2^{140}$ time steps (we have continued the simulation finding no periodicity for any reasonable computation time). 

\begin{figure} %Shown is a window where $t \in [0,150]$ and $i\in [1, 150]$
\includegraphics[width=0.49 \textwidth]{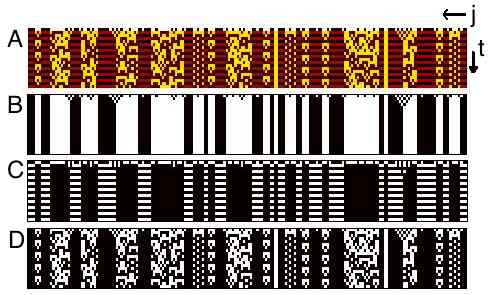}
\caption{\scriptsize{Detail of the spatiotemporal evolution of the CA dynamics for the first 20 time steps of panel C in Fig. \ref{chimepl}: Spatiotemporal evolution $x_{t}^{j}$ (A), $\mathbf{d}_{2}(0,x_{t}^{j})$ (B), $\mathbf{d}_{2}(0,x_{t}^{j}/2)$  (C) and $\mathbf{d}_{2}(0,x_{t}^{j}/4)$ (D). }} \label{dise}
\end{figure}

The dynamical behavior of the model is summarized in Fig. \ref{dise}. Panel A shows a detail of Fig. \ref{chimepl}C for the first 20 iteration steps. The spatiotemporal evolution of the layers $\mathbf{d}_{2}(0,x_{t}^{j}/2^{h})$, as obtained from Eqs. (\ref{layer0}) to (\ref{layer2}) is shown for $h=0$ (panel B), $h=1$ (panel C) and $h=2$ (panel D). We see that $x_{t}^{j}$ depends on the output of the majority rule (panel B) as explained above: \emph{If the output $y_{t}^{(0),j}$ of the majority rule dictated by Eq. (\ref{majomo}) is '0', $x_{t}^{j}$ is governed by Eq. (\ref{oscimo}); if the output $y_{t}^{(0),j}$ is '1', $x_{t}^{j}$ is found in the incoherent phase and takes a value governed by Eq. (\ref{ado})}.

There are $8^{8^{2\xi+1}}$ CAs in rule space with $p=8$ and range $2\xi+1$,
all described by Eq. (\ref{CAdyna}) or, equivalently, by \cite{semipredo}
\begin{equation}
x_{t+1}^{j}=\mathbf{d}_{8}\left(\sum_{k=-\xi}^{\xi}8^{k+\xi}x_{t}^{j+k} ,R \right) \label{theQUA}
\end{equation}
where $R \in [0,8^{8^{2\xi+1}}-1]$ (a non-negative integer) is the Wolfram code of the CA rule.
The CA model constructed in this paper belongs to this set and can be easily shown to have a huge Wolfram code located within the interval $8^{8^{2\xi+1}-2} < R < 8^{8^{2\xi+1}-1}$. Even for $\xi=1$, this is an enormous number. The general method presented in \cite{semipredo} (layer decomposition) and illustrated in this letter, makes it possible to systematically address such CAs in rule space.

In this article we have shown how a CA approximation can be constructed from any model of coupled phase oscillators. We have then presented a minimalistic CA model for chimera states and we have shown that they agree with a recent rigorous definition of chimeras \cite{ChimeraASH14, ChimeraASHW15b}. The main advantage of our model is that, owing to its simplicity, many features (domain size, transient duration, etc.) are estimated as a function of the only control parameter, the coupling range $\xi$. Under global coupling no chimera states of the kind described here are possible. Recently, chimera states under global coupling have been experimentally found in electrochemical systems \cite{ChimeraSCH14a} and modeled employing a modified complex Ginzburg-Landau equation \cite{ChimeraOrlov, ChimeraMiethe, ChimeraSCH14a} and Stuart-Landau oscillators \cite{ChimeraSCH14a, ChimeraSCHM15a, ChimeraSCHM15}. However, in these models the mechanism leading to the emergence of chimeras is different, since it is caused by the presence of a global constraint that introduces nontrivial correlations.  

\acknowledgments
Two anonymous referees are gratefully acknowledged for their helpful suggestions and stimulating remarks.

\bibliography{biblos}{}

\begin{thebibliography}{10}

\bibitem{ChimeraKUR02a}
Y.~Kuramoto and D.~Battogtokh,
\newblock Nonlin. Phen. in Complex Sys. {\bf 5}, 380 (2002).

\bibitem{ChimeraABR04}
D.~M. Abrams and S.~H. Strogatz,
\newblock Phys. Rev. Lett. {\bf 93}, 174102 (2004).

\bibitem{ChimeraTIN12}
M.~R. Tinsley, S.~Nkomo, and K.~Showalter,
\newblock Nature Phys. {\bf 8}, 662 (2012).

\bibitem{ChimeraHAG12}
A.~M. Hagerstrom {\em et~al.},
\newblock Nature Phys. {\bf 8}, 658 (2012).

\bibitem{ChimeraPAN15}
D.~M. Abrams and M.~J. Panaggio,
\newblock Nonlinearity {\bf 28}, R67 (2015).

\bibitem{ChimeraABR08}
D.~M. Abrams, R.~E. Mirollo, S.~H. Strogatz, and D.~A. Wiley,
\newblock Phys. Rev. Lett. {\bf 101}, 084103 (2008).

\bibitem{ChimeraSET08}
G.~C. Sethia, A.~Sen, and F.~M. Atay,
\newblock Phys. Rev. Lett. {\bf 100}, 144102 (2008).

\bibitem{ChimeraLAI09}
C.~R. Laing,
\newblock Physica D {\bf 238}, 1569 (2009).

\bibitem{ChimeraMOT10}
A.~E. Motter,
\newblock Nature Phys. {\bf 6}, 164 (2010).

\bibitem{ChimeraMAR10}
E.~A. Martens, C.~R. Laing, and S.~H. Strogatz,
\newblock Phys. Rev. Lett. {\bf 104}, 044101 (2010).

\bibitem{ChimeraOLM10}
S.~Olmi, A.~Politi, and A.~Torcini,
\newblock Europhys. Lett. {\bf 92}, 60007 (2010).

\bibitem{ChimeraBOR10}
G.~Bordyugov, A.~Pikovsky, and M.~Rosenblum,
\newblock Phys. Rev. E {\bf 82}, 035205 (2010).

\bibitem{ChimeraSHE10}
J.~H. Sheeba, V.~K. Chandrasekar, and M.~Lakshmanan,
\newblock Phys. Rev. E {\bf 81}, 046203 (2010).

\bibitem{ChimeraWOL11}
M.~Wolfrum and O.~E. Omelchenko,
\newblock Phys. Rev. E {\bf 84}, 015201 (2011).

\bibitem{ChimeraLAI11}
C.~R. Laing,
\newblock Physica D {\bf 240}, 1960 (2011).

\bibitem{ChimeraOME11}
I.~Omelchenko, Y.~Maistrenko, P.~H{\"o}vel, and E.~Sch{\"o}ll,
\newblock Phys. Rev. Lett. {\bf 106}, 234102 (2011).

\bibitem{ChimeraOME12}
I.~Omelchenko, B.~Riemenschneider, P.~H{\"o}vel, Y.~Maistrenko, and
  E.~Sch{\"o}ll,
\newblock Phys. Rev. E {\bf 85}, 026212 (2012).

\bibitem{ChimeraOME13}
I.~Omelchenko, O.~E. Omelchenko, P.~H{\"o}vel, and E.~Sch{\"o}ll,
\newblock Phys. Rev. Lett. {\bf 110}, 224101 (2013).

\bibitem{ChimeraNKO13}
S.~Nkomo, M.~R. Tinsley, and K.~Showalter,
\newblock Phys. Rev. Lett. {\bf 110}, 224102 (2013).

\bibitem{ChimeraHIZ13}
J.~Hizanidis, A.~Kanas, A.~Bezerianos, and T.~Bountis,
\newblock Int. J. Bifurcation Chaos {\bf 24}, 1450030 (2014).

\bibitem{ChimeraSET13}
G.~C. Sethia, A.~Sen, and G.~L. Johnston,
\newblock Phys. Rev. E {\bf 88}, 042917 (2013).

\bibitem{VGM3}
V.~Garc{\'\i}a-Morales,
\newblock Phys. Rev. E {\bf 88}, 042814 (2013).

\bibitem{ChimeraSET14}
G.~C. Sethia and A.~Sen,
\newblock Phys. Rev. Lett. {\bf 112}, 144101 (2014).

\bibitem{ChimeraYEL14}
A.~Yeldesbay, A.~Pikovsky, and M.~Rosenblum,
\newblock Phys. Rev. Lett. {\bf 112}, 144103 (2014).

\bibitem{ChimeraBOE15}
F.~B{\"o}hm, A.~Zakharova, E.~Sch{\"o}ll, and K.~L{\"u}dge,
\newblock Phys. Rev. E {\bf 91}, 040901R (2015).

\bibitem{ChimeraBUS15}
A.~Buscarino, M.~Frasca, L.~V. Gambuzza, and P.~H{\"o}vel,
\newblock Phys. Rev. E {\bf 91}, 022817 (2015).

\bibitem{ChimeraOME15}
I.~Omelchenko, A.~Provata, J.~Hizanidis, E.~Sch{\"o}ll, and P.~H{\"o}vel,
\newblock Phys. Rev. E {\bf 91}, 022917 (2015).

\bibitem{ChimeraOME15A}
I.~Omelchenko, A.~Zakharova, P.~H{\"o}vel, J.~Siebert, and E.~Sch{\"o}ll,
\newblock Chaos {\bf 25}, 083104 (2015).

\bibitem{ChimeraBAN15}
P.~S. Dutta and T.~Banerjee,
\newblock Phys. Rev. E {\bf 92}, 042919 (2015).

\bibitem{ChimeraASH14}
P.~Ashwin and O.~Burylko,
\newblock Chaos {\bf 25}, 013106 (2015).

\bibitem{ChimeraASHW15b}
C.~Bick and P.~Ashwin,
\newblock Nonlinearity {\bf 29}, 1468 (2016).

\bibitem{ChimeraHIZ15}
J.~Hizanidis {\em et~al.},
\newblock Phys. Rev. E {\bf 92}, 012915 (2015).

\bibitem{ChimeraSCHM15a}
L.~Schmidt and K.~Krischer,
\newblock Phys. Rev. Lett. {\bf 114}, 034101 (2015).

\bibitem{ChimeraSCHM15}
L.~Schmidt and K.~Krischer,
\newblock Chaos {\bf 25}, 064401 (2015).

\bibitem{ChimeraHAU15}
S.~W. Haugland, L.~Schmidt, and K.~Krischer,
\newblock Sci. Rep. {\bf 5}, 9883 (2015).

\bibitem{ChimeraKEM16}
F.~P. Kemeth, S.~W. Haugland, L.~Schmidt, I.~G. Kevrekidis, and K.~Krischer,
\newblock arXiv:1603.01110 [nlin.CD]  (2016).

\bibitem{ChimeraMAR13}
E.~A. Martens, S.~Thutupalli, A.~Fourriere, and O.~Hallatschek,
\newblock Proc. Natl. Acad. Sci. USA {\bf 110}, 10563 (2013).

\bibitem{ChimeraLAR13}
L.~Larger, B.~Penkovsky, and Y.~Maistrenko,
\newblock Phys. Rev. Lett. {\bf 111}, 054103 (2013).

\bibitem{ChimeraKAP14}
T.~Kapitaniak, P.~Kuzma, J.~Wojewoda, K.~Czolczynski, and Y.~Maistrenko,
\newblock Sci. Rep. {\bf 4}, 6379 (2014).

\bibitem{ChimeraWIC13}
M.~Wickramasinghe and I.~Z. Kiss,
\newblock PLoS ONE {\bf 8}, e80586 (2013).

\bibitem{ChimeraWIC14}
M.~Wickramasinghe and I.~Z. Kiss,
\newblock Phys. Chem. Chem. Phys. {\bf 16}, 18360 (2014).

\bibitem{ChimeraSCH14a}
L.~Schmidt, K.~Sch{\"o}nleber, K.~Krischer, and V.~Garcia-Morales,
\newblock Chaos {\bf 24}, 013102 (2014).

\bibitem{ChimeraGAM14}
L.~V. Gambuzza {\em et~al.},
\newblock Phys. Rev. E {\bf 90}, 032905 (2014).

\bibitem{ChimeraROS14a}
D.~P. Rosin, D.~Rontani, N.~D. Haynes, E.~Sch{\"o}ll, and D.~J. Gauthier,
\newblock Phys. Rev. E {\bf 90}, 030902R (2014).

\bibitem{ChimeraLAR15}
L.~Larger, B.~Penkovsky, and Y.~Maistrenko,
\newblock Nature Commun. {\bf 6}, 7752 (2015).

\bibitem{ChimeraGON14}
J.~C. Gonzalez-Avella, M.~Cosenza, and M.~S. Miguel,
\newblock Physica A {\bf 399}, 24 (2014).

\bibitem{ChimeraMOT13a}
A.~E. Motter, S.~A. Myers, M.~Anghel, and T.~Nishikawa,
\newblock Nature Phys. {\bf 9}, 191 (2013).

\bibitem{ChimeraROT14}
A.~Rothkegel and K.~Lehnertz,
\newblock New J. Phys. {\bf 16}, 055006 (2014).

\bibitem{ChimeraRAT00}
N.~Rattenborg, C.~J. Amlaner, and S.~L. Lima,
\newblock Neurosci. Biobehav. Rev. {\bf 24}, 817 (2000).

\bibitem{Wolfram}
S.~Wolfram,
\newblock {\em A New Kind of Science} (Wolfram Media Inc., Champaign, IL,
  2002).

\bibitem{Ilachinski}
A.~Ilachinski,
\newblock {\em Cellular Automata: a Discrete Universe} (World Scientific,
  Singapore, 2001).

\bibitem{Adamatzky}
A.~Adamatzky,
\newblock {\em Identification of Cellular Automata} (Taylor and Francis,
  London, 1994).

\bibitem{McIntosh}
H.~V. McIntosh,
\newblock {\em One Dimensional Cellular Automata} (Luniver Press, Frome, UK,
  2009).

\bibitem{Wuensche}
A.~Wuensche and M.~Lesser,
\newblock {\em The Global Dynamics of Cellular Automata} (Addison-Wesley,
  Reading, MA, 1992).

\bibitem{Ceccherini}
T.~Ceccherini-Silberstein and M.~Coornaert,
\newblock {\em Cellular Automata and Groups} (Springer, Berlin, 2010).

\bibitem{VGM1}
V.~Garc{\'\i}a-Morales,
\newblock Phys. Lett. A {\bf 376}, 2645 (2012).

\bibitem{VGM2}
V.~Garc{\'\i}a-Morales,
\newblock Phys. Lett. A {\bf 377}, 276 (2013).

\bibitem{ChimeraMAK}
S.~D. Makovetskiy and D.~Makovetskii,
\newblock arXiv:cond-mat/0410460v2 [cond-mat.other]  (2005).

\bibitem{JPHYSA}
V.~Garc{\'\i}a-Morales,
\newblock arXiv:1602.00289 [nlin.CG]  (2016).

\bibitem{Omohundro}
S.~Omohundro,
\newblock Physica D {\bf 10}, 128 (1984).

\bibitem{Lind}
D.~Lind and B.~Marcus,
\newblock {\em Symbolic Dynamics and Coding} (Cambridge University Press,
  Cambridge, UK, 1995).

\bibitem{CHAOSOLFRAC}
V.~Garc{\'\i}a-Morales,
\newblock Chaos Sol. Fract. {\bf 83}, 27 (2016), cond-mat/1505.02547v3.

\bibitem{PHYSAFRAC}
V.~Garc{\'\i}a-Morales,
\newblock Physica A {\bf 447}, 535 (2016), cs.OH/1507.01444v3.

\bibitem{semipredo}
V.~Garc{\'\i}a-Morales,
\newblock Commun. Nonlinear Sci. Numer. Simulat. {\bf 39}, 81 (2016).

\bibitem{Andrews}
G.~E. Andrews,
\newblock {\em Number Theory} (Dover, New York, NY, 1994).

\bibitem{Vichniac}
G.~Vichniac,
\newblock Physica D {\bf 10}, 96 (1984).

\bibitem{Tchuente}
M.~Tchuente,
\newblock {\em Contribution a l'etude des methodes de calcul pour des systemes
  de type cooperatif} (Thesis, University of Grenoble, France, 1982).

\bibitem{Goles}
E.~Goles and M.~Tchuente,
\newblock Disc. Appl. Math. {\bf 8}, 319 (1984).

\bibitem{GolesBOOK}
E.~Goles and S.~Martinez,
\newblock {\em Neural and Automata Networks} (Kluwer, Amsterdam, 1990).

\bibitem{Agur}
Z.~Agur,
\newblock Complex Systems {\bf 5}, 351 (1991).

\bibitem{Agur2}
Z.~Agur,
\newblock IMA J. Math. Appl. Med. Biol. {\bf 4}, 295 (1987).

\bibitem{Dolnik}
L.~Yang, M.~Dolnik, A.~M. Zhabotinsky, and I.~R. Epstein,
\newblock Phys. Rev. E {\bf 62}, 6414 (2000).

\bibitem{KrischerBaba}
N.~Baba and K.~Krischer,
\newblock Chaos {\bf 18}, 015103 (2008).

\bibitem{Wolfram2}
S.~Wolfram,
\newblock {\em Cellular Automata and Complexity: Collected Papers}
  (Addison-Wesley, Reading, MA, 1994).

\bibitem{Chate2}
H.~Chate and P.~Manneville,
\newblock J. Stat. Phys. {\bf 56}, 357 (1989).

\bibitem{Chate3}
H.~Chate and P.~Manneville,
\newblock Physica D {\bf 45}, 122 (1990).

\bibitem{KuramotoBOOK}
Y.~Kuramoto,
\newblock {\em Chemical Oscillations, Waves and Turbulence} (Springer, New
  York, 1984).

\bibitem{Aranson}
I.~S. Aranson and L.~Kramer,
\newblock Rev. Mod. Phys. {\bf 74}, 99 (2002).

\bibitem{contemphys}
V.~Garc{\'\i}a-Morales and K.~Krischer,
\newblock Contemp. Phys. {\bf 53}, 79 (2012).

\bibitem{ChimeraOrlov}
V.~Garc\'{\i}a-Morales, A.~Orlov, and K.~Krischer,
\newblock Phys. Rev. E {\bf 82}, 065202(R) (2010).

\bibitem{ChimeraMiethe}
I.~Miethe, V.~Garc\'{\i}a-Morales, and K.~Krischer,
\newblock Phys. Rev. Lett. {\bf 102}, 194101 (2009).

\end{thebibliography}
\bibliographystyle{h-physrev3.bst}
\end{document}